\documentclass[a4paper, amsfonts, amssymb, amsmath, reprint, showkeys, nofootinbib, twoside,notitlepage,onecolumn]{revtex4-1}

\usepackage{amsmath,amstext}
\usepackage[T1]{fontenc}
\usepackage{amssymb}
\usepackage{graphicx}
\usepackage{ae,aecompl}

\DeclareFontFamily{OT1}{pzc}{}
\DeclareFontShape{OT1}{pzc}{m}{it}{<-> s * [1.10] pzcmi7t}{}
\DeclareMathAlphabet{\mathpzc}{OT1}{pzc}{m}{it}

\usepackage{hyperref}
\usepackage{amsmath}
\usepackage{amssymb}
\usepackage{mathtools}
\usepackage{bm}
\usepackage{cleveref}
\usepackage{tensor}
\usepackage{braket}
\usepackage{enumitem}
\usepackage{mhchem}
\usepackage{amsthm}
\usepackage{nccmath}
\usepackage{mathrsfs}
\usepackage{color}

\def\be{\begin{equation}}
\def\ee{\end{equation}}
\def\beq{\begin{eqnarray}}
\def\eeq{\end{eqnarray}}

\theoremstyle{definition}

\theoremstyle{theorem}

\begin{document}
\title{Thermal boundaries for relativistic fluids without a conserved charge}
\author{L.~Gavassino$^1$, S.~Schlichting$^2$, G.S.~Denicol$^3$}
\affiliation{$^1$Department of Applied Mathematics and Theoretical Physics, University of Cambridge, Wilberforce Road, Cambridge CB3 0WA, United Kingdom\\
$^2$Fakult\"{a}t f\"{u}r Physik, Universit\"{a}t Bielefeld, D-33615 Bielefeld, Germany\\
$^3$Instituto de Física, Universidade Federal Fluminense Av. Gal. Milton Tavares de Souza, S/N, 24210-346, Gragoatá, Niterói, Rio de Janeiro, Brazil}

\begin{abstract}
We study an ultrarelativistic fluid with no conserved particle number in a planar slab geometry bounded by two parallel, thermally conducting plates held at fixed (possibly different) temperatures. The fluid-wall interaction is described within kinetic theory by modeling the fluid as self-interacting radiation: a gas of bosons emitted and absorbed by the walls according to black-body laws, while mutual collisions enforce local equilibration.
From this microscopic setup we derive effective boundary conditions for the hydrodynamic fields, finding that the walls behave as modified absorbing boundaries (and not as thermostats). In particular, the fluid temperature at the boundary does not generally coincide with the wall temperature. We prove that the resulting initial-boundary-value problem is well posed. When the plates are held at different temperatures, the system develops a uniform steady flow from the hotter to the colder wall, with energy density equal to the arithmetic mean of the corresponding black-body energy densities in the limit of a small temperature difference.
\end{abstract} 
\maketitle

\vspace{-1.1cm}
\section{Introduction}
\vspace{-0.4cm}

Heat conduction is one of the most fundamental nonequilibrium processes in physics. A paradigmatic setting consists of a fluid confined between walls maintained at different temperatures, where the transfer of energy between the fluid and the reservoirs is described through boundary conditions imposed at the fluid-wall interface. In nonrelativistic hydrodynamics, rigid walls at rest are commonly modeled as Dirichlet-type boundaries: both the normal and tangential components of the fluid velocity vanish at the interface (no penetration and no slip), and if the wall is thermally conducting, the fluid temperature is assumed to coincide with the wall temperature at the point of contact. When a conserved particle current is present, this prescription also admits a natural relativistic interpretation in the Eckart frame, where the fluid velocity is identified with the particle flux \cite{GavassinoCouettetype:2025kad}. 

In other hydrodynamic frames, however, its physical meaning becomes less transparent. For example, in the Landau frame, the velocity $u^\mu$ is defined by the energy current rather than by the particle flux. Consequently, whenever heat is exchanged between the fluid and the wall, a nonvanishing normal component of $u^\mu$ generally arises, reflecting the transfer of energy across the interface. The ambiguity is even more pronounced within the BDNK formalism \cite{Bemfica2019_conformal1,Kovtun2019,BemficaDNDefinitivo2020}, where all hydrodynamic fields are subject to frame redefinitions and even the notion of temperature is not unique. As a result, the interpretation of Dirichlet boundary conditions for the temperature becomes intrinsically frame-dependent.

The absence of a conserved particle current introduces an additional challenge, affecting both the hydrodynamic and thermodynamic descriptions of heat transfer. From the hydrodynamic perspective, the Eckart frame is no longer available, so the conventional boundary conditions no longer admit a clear physical interpretation. From the thermodynamic perspective, the constitutive description of heat flow also changes. In the presence of a conserved charge, the heat current is driven by thermodynamic forces involving gradients of chemical potential over temperature, $\sim\nabla(\mu/T)$ \cite{landau6}. At vanishing chemical potential, this thermodynamic force vanishes identically, calling into question the conventional thermodynamic description of heat transfer between reservoirs held at different temperatures.

The implications of these observations become apparent in the simplest heat-conduction problem: a fluid at zero chemical potential confined between two parallel plates held at different temperatures. Since the equation of state depends only on the temperature, $P=P(T)$, imposing the usual Dirichlet boundary conditions necessarily generates a pressure gradient across the slab. Under stationary planar symmetry, however, such a pressure gradient cannot be sustained by a regular flow: as shown in Appendix \ref{aaa}, the only stationary solution is a shock profile in which the fluid is emitted and absorbed by the walls at nearly sonic speed. This behavior is clearly incompatible with the expected picture of steady heat conduction, indicating that the interaction between the fluid and the thermal reservoirs cannot be captured by the standard Dirichlet prescription.

In this work, we derive physically motivated boundary conditions for ultrarelativistic fluids at zero chemical potential in contact with thermally conducting walls. Our approach is based on kinetic theory, describing the fluid as a gas of interacting massless bosons that exchange particles with the walls through black-body emission and absorption processes. The resulting effective boundary conditions not only have a clear physical interpretation but also render the corresponding initial-boundary-value problem well posed. This microscopic picture naturally gives rise to a stationary conducting state, in which particles emitted by the hotter wall diffuse through the fluid toward the colder reservoir while collisions maintain local thermal equilibrium. As a result, the system supports a constant heat flux between the walls, while the energy density remains spatially uniform and \textit{equal to the arithmetic mean} of the corresponding black-body energy densities.

Throughout the article, we work in Minkowski space, with metric signature $(-,+,+,+)$, and work in natural units, with $c=\hbar=k_B=1$. Greek indices run from $0$ to $3$, while Latin indices run from $1$ to $3$.

\vspace{-0.2cm}
\section{Kinetic framework}
\vspace{-0.2cm}

In this section we introduce a kinetic model for interacting, massless, nonconserved bosons, which will later provide the microscopic underpinning for our boundary-data prescription.

\subsection{The Boltzmann equation}

Let $f(x^\mu,p^\alpha)$ denote the single-particle distribution function of the boson gas, giving the occupation number of one-boson states with four-momentum $p^\alpha$ ($p^\alpha p_\alpha=0$) at the spacetime point $x^\mu$. Treating the wall as a continuous material medium occupying a spacetime region $\mathcal{W}$ and interacting with the bosons, the evolution of $f$ is governed by the relativistic Boltzmann equation
\begin{equation}\label{Boltzmann}
p^\mu \partial_\mu f=\mathcal{C}_{bb}+\mathcal{C}_{bw}\, .
\end{equation}
Here $\mathcal{C}_{bb}[f]$ denotes a particle-nonconserving collision integral describing interactions among the bosons, while $\mathcal{C}_{bw}[f,\text{``wall parameters''}]$ represents absorption and emission processes (we neglect scattering) carried out by the wall itself within $\mathcal{W}$. We will assume throughout that the walls are spatially uniform, and in local thermal equilibrium, with grey opacity. Under these conditions, the Kirchhoff-Planck relation \cite{mihalas_book} implies
\begin{equation}\label{wallcollisions}
\frac{1}{U_\mu p^\mu}\,\mathcal{C}_{bw}=
\begin{cases}
0, & \text{outside }\mathcal{W}\,,\\[6pt]
\dfrac{f-f_{eq}(T,U_\mu p^\mu)}{\tau}, & \text{inside }\mathcal{W}\, .
\end{cases}
\end{equation}
Here, $T=\mathrm{const}>0$ is the wall's temperature, $U^\mu$ is the wall's velocity, and $\tau=\mathrm{const}>0$ is the absorption length. The first term in parentheses is the Black-Body (equivalently, Bose-Einstein) distribution $f_{eq}=(e^{-U_\mu p^\mu/T}-1)^{-1}$ that the bosons would exhibit if they were in local equilibrium with the wall. For the remainder of this paper, we assume that the walls are at rest, $U^\mu = (1,0,0,0)$. 

We note that the collision term in \cref{wallcollisions} is reminiscent of the Anderson-Witting relaxation-time approximation \cite{AndersonWitting1974}, with the crucial distinction that the reference equilibrium distribution $f_{eq}$ is not fixed by Landau matching but instead imposed by the wall. As we shall see, when the fluid undergoes collective motion relative to the wall, $\mathcal{C}_{bw}$ acts as an effective friction term, whereas temperature differences between the fluid and the wall give rise to heating or cooling.

\subsection{Energy-momentum balance}

The stress-energy tensor of the fluid reads
\begin{equation}
T^{\mu \nu}=\int \dfrac{g\, d^3p}{(2\pi)^3p^0} \, p^\mu p^\nu f \, ,
\end{equation}
where $g\,{\in}\, \mathbb{N}$ is a spin/color degeneracy factor. The interactions among the bosons (represented by $\mathcal{C}_{bb}$) always conserve energy-momentum. Hence, when we integrate both sides of \eqref{Boltzmann} in $p^\nu d^3 p/p^0$, we obtain the following balance laws:
\begin{equation}\label{balanceenergy}
\partial_\mu T^{\mu 0}=
\begin{cases}
0, & \text{outside }\mathcal{W}\,,\\[6pt]
\dfrac{aT^4-T^{00}}{\tau}, & \text{inside }\mathcal{W}\, ,
\end{cases}
\end{equation}
for the energy (with $a=g\pi^2/30$ the radiation constant), and
\begin{equation}\label{balancemomentum}
\partial_\mu T^{\mu k}=
\begin{cases}
0, & \text{outside }\mathcal{W}\,,\\[6pt]
-\dfrac{T^{0k}}{\tau}, & \text{inside }\mathcal{W}\, ,
\end{cases}
\end{equation}
for the momentum. Note that the above equations are exact, within the model assumptions.

We see from \eqref{balanceenergy} and \eqref{balancemomentum} that absorption and emission by the wall generate an effective four-force density acting on the boson gas, with the wall experiencing the equal and opposite reaction. This reaction has a clear physical interpretation: its time component represents the net power density absorbed (i.e., energy exchanged per unit time and volume) by the wall, while its spatial components correspond to the force density associated with momentum transfer due to radiation pressure. In the following, we assume that the walls have infinite heat capacity and mass, so that they neither accelerate nor change temperature.

\vspace{-0.3cm}
\subsection{The non-self-interacting limit}\label{nonself}
\vspace{-0.2cm}

Let us verify explicitly that, in the limit $\mathcal{C}_{bb}\to 0$ (i.e., in the absence of boson-boson interactions), the model reproduces the standard results of radiative transfer theory. Specifically, consider two semi-infinite planar walls: a left wall occupying $x<-L/2$ at temperature $T_\ell$, and a right wall occupying $x>L/2$ at temperature $T_{\mathit r}$. In this setting, the energy flux exchanged between the walls is expected to obey the Stefan-Boltzmann law,
\begin{equation}\label{stefan}
\mathcal{F}=\frac{a}{4}\left(T_\ell^4-T_{\mathit r}^4\right).
\end{equation}
We now show that this result follows directly from the kinetic description.

Under stationary conditions and for the geometry described above, equation \eqref{Boltzmann} reduces to
\begin{equation}\label{BoltzmannStefan}
\frac{p^x}{p^0}\,\partial_x f=
\begin{cases}
-\dfrac{f-f_{eq}(T_\ell)}{\tau_\ell}, & x\le -L/2\,,\\[6pt]
0, & -L/2<x<L/2\,,\\[6pt]
-\dfrac{f-f_{eq}(T_{\mathit r})}{\tau_{\mathit r}}, & x\ge L/2\, .
\end{cases}
\end{equation}
It follows immediately that $f$ is constant in the region between the walls. Moreover, for $p^x>0$, solving the Boltzmann equation inside the left wall yields solutions that diverge as $x\to -\infty$ unless one selects $f=f_{eq}(T_\ell)$. Similarly, for $p^x<0$, boundedness as $x\to +\infty$ requires $f=f_{eq}(T_{\mathit r})$ inside the right wall. Consequently, in the vacuum region between the plates one finds
$f=f_{eq}(T_\ell)\Theta(p^x)+f_{eq}(T_{\mathit r})\Theta(-p^x)$ by continuity,
i.e. particles emerging from each wall follow the black-body distribution associated with that wall.
The resulting stress-energy tensor takes the form
\begin{equation}\label{emomentumstefan}
T^{\mu\nu}=a
\begin{bmatrix}
\dfrac{T_\ell^4+T_{\mathit r}^4}{2} & \dfrac{T_\ell^4-T_{\mathit r}^4}{4} & 0 & 0 \\
\dfrac{T_\ell^4-T_{\mathit r}^4}{4} & \dfrac{T_\ell^4+T_{\mathit r}^4}{6} & 0 & 0 \\
0 & 0 & \dfrac{T_\ell^4+T_{\mathit r}^4}{6} & 0 \\
0 & 0 & 0 & \dfrac{T_\ell^4+T_{\mathit r}^4}{6}
\end{bmatrix},
\end{equation}
so that $T^{0x}=\mathcal{F}$ indeed reproduces the Stefan-Boltzmann law~\eqref{stefan}.

The key point of this analysis is that the same result could have been obtained by directly imposing, as boundary data, that particles emitted by each wall follow the corresponding black-body distribution. Here, however, this condition emerges dynamically by integrating equation \eqref{BoltzmannStefan} inside the walls and requiring boundedness of $f$ at spatial infinity. As we shall see, this latter procedure extends straightforwardly (under the same geometry) to the interacting case, whereas the former does not.

\vspace{-0.2cm}
\subsection{The fluid limit}
\vspace{-0.2cm}
We are interested in the regime in which relaxation driven by $\mathcal{C}_{bb}$ occurs on parametrically shorter timescales than those associated with the interaction with the walls. In this limit, boson-boson collisions maintain local thermodynamic equilibrium throughout the fluid, although the fluid itself need not be in equilibrium with the walls. The system may therefore be described by relativistic hydrodynamics. Throughout this work, we further restrict attention to situations in which the fluid velocity relative to the walls remains small. This assumption is well satisfied in the stationary configurations studied below, where the energy flux is much smaller than the energy density. Expanding the stress-energy tensor to first order in the spatial components of the four-velocity $u^\mu$, it is convenient to express it directly in terms of the conserved variables,
\newpage
\begin{equation}\label{idealfluid}
T^{\mu \nu}=
\begin{bmatrix}
\varepsilon & \mathcal{F}^x & \mathcal{F}^y & \mathcal{F}^z \\
\mathcal{F}^x & \varepsilon/3 & 0 & 0 \\
\mathcal{F}^y & 0 & \varepsilon/3 & 0 \\
\mathcal{F}^z & 0 & 0 & \varepsilon/3
\end{bmatrix}
+\mathcal{O}\!\left(\left[u^k\right]^2\right),
\end{equation}
where, to this order, the Landau velocity can be expressed in the simple form $u^k=3\mathcal{F}^k/(4\varepsilon)$.

To account for departures from local equilibrium, we include the leading dissipative correction through the Navier--Stokes constitutive relation,
\begin{equation}
\pi_{jk}
=
-2\eta
\left(
\partial_{(j}u_{k)}
-\frac13\delta_{jk}\partial_lu^l
\right),
\end{equation}
where $\eta$ denotes the shear viscosity. Although the interaction with the walls provides an additional mechanism for relaxing momentum, we assume that the corresponding absorption timescales are much longer than the microscopic relaxation time associated with boson-boson collisions. Consequently, the non-equilibrium distribution generated by velocity gradients relaxes to its Navier--Stokes form long before the walls appreciably absorb momentum. The microscopic processes responsible for shear viscosity are therefore unaffected by the presence of the walls, and the same transport coefficient $\eta$ may be used both inside and outside the wall regions.

Finally, it is convenient to express the viscous stress directly in terms of the conserved variables. Since $u^k=\mathcal{O}(\mathcal{F})$, the small-velocity expansion further implies that $|\partial_tu_k|
\sim
|\partial_ku_j|
\sim
|\partial_k\varepsilon|
=
\mathcal{O}(u)$,
so that nonlinear products such as $\mathcal{F}_k\partial_j\varepsilon$ are of second order in the expansion and may consistently be neglected. Substituting the linear relation between $u^k$ and $\mathcal{F}^k$ into the constitutive equation then yields
\begin{equation}\label{NavierStokesSress}
\pi_{jk}
=
-2D
\left(
\partial_{(j}\mathcal{F}_{k)}
-\frac13\delta_{jk}\partial_l\mathcal{F}^l
\right),
\end{equation}
where we introduced the transport coefficient $D=3\eta/(4\varepsilon)$,
which can be identified as the momentum diffusivity associated with shear waves. Throughout the remainder of this work, $D$ will be treated as a constant for simplicity.

\section{Transversal flows}

Let us consider the following warm-up question: In the limit where the absorption length $\tau$ of the wall is very small, do tangential flows obey the no-slip boundary condition? To answer this question, we work in the following slab geometry: a left wall occupies the region $x<-L/2$ and a right wall occupies the region $x>L/2$, both with velocity $U^\mu=(1,0,0,0)$, temperature $T$, and absorption length $\tau$. We then consider flows pointing in the $y$ direction,
$\mathcal{F}^k=(0,\mathcal{F},0)$, so the fluid motion is transversal to the slab geometry (we also set $\varepsilon=\text{const}$). In the ideal-fluid case, the resulting dynamics is trivial in the gap between the walls, because there is no friction between fluid layers, and any profile $\mathcal{F}(x)$ is a stationary solution of the hydrodynamic equations. Hence, we must introduce a shear viscosity.

\subsection{Mathematical setup}

In the slab geometry considered here, the only nontrivial balance equation is the $y$ component of the momentum conservation law. Moreover, the only viscous contribution that enters is $\pi_{xy} = -D \partial_x \mathcal{F}$. The evolution of $\mathcal{F}$ is therefore governed by
\begin{equation}\label{transversalShear}
\dfrac{1}{D}\partial_t \mathcal{F} - \partial_x^2 \mathcal{F}
=
- \frac{\mathcal{F}}{D\tau}
\begin{cases}
1, & x \le -L/2\,,\\[2pt]
0, & -L/2 < x < L/2\,,\\[2pt]
1, & x \ge L/2\, .
\end{cases}
\end{equation}
This equation is formally equivalent to a one-dimensional Schr\"odinger equation in imaginary time, with an external potential given by a symmetric square barrier of height $1/(D\tau)$. Under the assumption that $\mathcal{F}(\pm \infty)=0$, the solution may therefore be constructed by expanding $\mathcal{F}$ in eigenmodes of the operator $-\partial_x^2 + (D\tau)^{-1}\Theta(|x|-L/2)$. Since this operator is self-adjoint and positive definite, its spectrum is real and positive, and the corresponding evolution is purely dissipative. The spectrum consists of a discrete sector, associated with eigenvalues below $1/(D\tau)$, and a continuous sector, corresponding to eigenvalues above $1/(D\tau)$. In the limit $D\tau \ll L^2$, which is the relevant regime of the effective boundary-value problem (recall we plan to take the limit $\tau \to 0$), the continuous modes become infinitely damped. We shall therefore restrict attention to the discrete part of the spectrum.

\subsection{Eigensolutions}

By symmetry, the eigenfunctions may be chosen to have definite parity. Here, we restrict attention to the even sector, the odd case being entirely analogous. Limiting ourselves to the discrete branch of the spectrum, the basis of eigensolutions is
\begin{equation}\label{modesDiffusia}
\mathcal{F}(t,x) 
=
e^{-Dk^2 t}
\begin{cases}
\cos\!\left(k\frac{L}{2}\right)
\exp\!\left[\sqrt{\frac{1}{D\tau}-k^2}\,\left(x+\frac{L}{2}\right)\right], 
& x \le -L/2\,,\\[4pt]
\cos(kx), 
& -L/2 < x < L/2\,,\\[4pt]
\cos\!\left(k\frac{L}{2}\right)
\exp\!\left[-\sqrt{\frac{1}{D\tau}-k^2}\,\left(x-\frac{L}{2}\right)\right], 
& x \ge L/2\, ,
\end{cases}   
\end{equation}
where $k\in\mathbb{R}$ (as required by positivity of the spectrum) and $D\tau k^2<1$ (since we are selecting discrete eigenvalues).
The allowed values of $k$ are fixed by imposing continuity of $\partial_x\mathcal{F}$ at $x=\pm L/2$, so that \eqref{transversalShear} is satisfied in the distributional sense. This matching condition yields the transcendental equation
\begin{equation}\label{shieddnoz}
\cos^2\!\left(k\tfrac{L}{2}\right)=D\tau k^2\, ,
\end{equation}
which may be solved numerically or graphically (as is done in quantum mechanics textbooks). 

\subsection{Boundary conditions in the limit of vanishing absorption length}

Let us now consider the limit $D\tau \ll L^2$, in which the height of the barrier diverges. By analogy with ordinary quantum mechanics, one expects $\mathcal{F}(\pm L/2)$ to vanish in this limit, as a wavefunction would in the presence of an infinite potential barrier. This behavior is indeed manifest in equation \eqref{shieddnoz}: as $D\tau/L^2 \to 0$, the modes \eqref{modesDiffusia} become exponentially suppressed outside the gap and are effectively confined to $-L/2<x<L/2$, vanishing at the interfaces. The quantization condition reduces to the standard Dirichlet one, $kL\in \pi(1+2\mathbb{N})$.
In this regime, the dynamics reduces to the diffusion equation on $(-L/2,L/2)$ supplemented by no-slip (i.e. Dirichlet) boundary conditions,
\begin{equation}\label{dirichl}
\mathcal{F}(\pm L/2)=0 \, .
\end{equation}
The resulting initial-boundary value problem is well-posed. Existence of solutions follows by expanding the initial data in the Fourier basis formed by the modes \eqref{modesDiffusia} (together with the analogous sine modes) satisfying $\cos(k\frac{L}{2})=0$. Uniqueness and continuous dependence on the initial data follow from standard energy estimates. Defining
\begin{equation}
E(t)=\dfrac{1}{2}\int_{-L/2}^{L/2} \mathcal{F}^2 \, dx,
\end{equation}
and using \eqref{dirichl}, one finds
$\Dot{E}=
- D \int (\partial_x\mathcal{F})^2 dx
\le 0 $. Hence, $E(0)=0$ implies $E(t)=0$ (uniqueness), while $E(t)\le E(0)$ ensures continuous dependence on the initial data.

In summary: Transversal flows of a viscous fluid in contact with thermally conducting walls do obey no-slip boundary conditions, and their evolution defines a well-posed initial-boundary value problem.

\section{Longitudinal flows}\label{invishidSec}
\vspace{-0.3cm}

We now investigate energy transport in the hydrodynamic regime between two plates held at different temperatures, restricting attention to flows along the $x$ direction, so that $\mathcal{F}^k=(\mathcal{F},0,0)$. Throughout this article, we consider the same geometry introduced in Sec.~\ref{nonself}: a left wall occupying $x<-L/2$, characterized by temperature $T_\ell$ and absorption length $\tau_\ell$, and a right wall occupying $x>L/2$, characterized by temperature $T_{\mathit r}$ and absorption length $\tau_{\mathit r}$.

Our strategy is to solve the hydrodynamic balance laws throughout the entire domain, including the interior of the walls, subject to the requirement that the fields asymptotically approach the corresponding black-body equilibrium states. Specifically, we impose
\begin{equation}
\begin{split}
&\varepsilon\rightarrow aT_\ell^4, \,\,\,\,\,
\mathcal{F}\rightarrow0, \,\,\,\,\,
\text{as }x\rightarrow-\infty,\\
&\varepsilon\rightarrow aT_{\mathit r}^4, \,\,\,\,\,
\mathcal{F}\rightarrow0,
\,\,\,\,\,
\text{as }x\rightarrow+\infty.\\
\end{split}
\end{equation}
Restricting the resulting solution to the region $-L/2<x<L/2$ between the plates then determines the effective boundary conditions for the hydrodynamic fields.

For a slab geometry, the balance laws \eqref{balanceenergy}-\eqref{balancemomentum} in the fluid limit reduce to
\vspace{-0.1cm}
\begin{equation}\label{ViscosoneLong}
\partial_t \varepsilon+\partial_x \mathcal{F}=
\begin{cases}
\dfrac{aT_\ell^4-\varepsilon}{\tau_\ell}, & x\le -L/2\,,\\[4pt]
0, & -L/2<x<L/2\,,\\[4pt]
\dfrac{aT_{\mathit r}^4-\varepsilon}{\tau_{\mathit r}}, & x\ge L/2\, ,
\end{cases}
\qquad
\partial_t \mathcal{F}+\frac{1}{3}\partial_x \varepsilon-\dfrac{4}{3}D\partial^2_x \mathcal{F}=
\begin{cases}
-\dfrac{\mathcal{F}}{\tau_\ell}, & x\le -L/2\,,\\[4pt]
0, & -L/2<x<L/2\,,\\[4pt]
-\dfrac{\mathcal{F}}{\tau_{\mathit r}}, & x\ge L/2\, .
\end{cases}
\end{equation}
The first equation expresses local energy conservation supplemented by the exchange of energy with the walls through black-body emission and absorption. The second equation describes momentum conservation, where the additional diffusive term proportional to $D$ accounts for viscous transport via, $\pi_{xx} \approx-\frac{4}{3}D\,\partial_x\mathcal{F}$. Together, these equations form a linear inhomogeneous system with piecewise constant source terms. Consequently, every solution approaching the prescribed equilibrium states at spatial infinity may be decomposed as
\begin{equation}
\begin{split}
\varepsilon(t,x) &
=
\varepsilon_0(x)
+
\Delta\varepsilon(t,x),\\
\mathcal{F}(t,x) &
=
\mathcal{F}_0(x)
+
\Delta\mathcal{F}(t,x),\\
\end{split}
\end{equation}
where $\{\varepsilon_0,\mathcal{F}_0\}$ denotes the stationary solutions, while $\{\Delta\varepsilon,\Delta\mathcal{F}\}$ satisfies the associated homogeneous problem. We first determine the stationary solution and subsequently investigate the relaxation of fluctuations around it.

\vspace{-0.5cm}
\subsection{The stationary solution}
\vspace{-0.4cm}

\begin{figure}[h!]
    \centering
\includegraphics[width=0.44\linewidth]{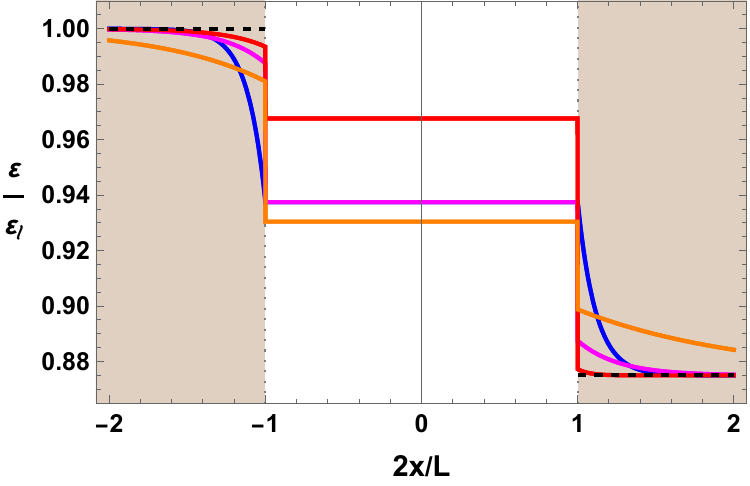}\hspace{0.05\linewidth}
\includegraphics[width=0.44\linewidth]{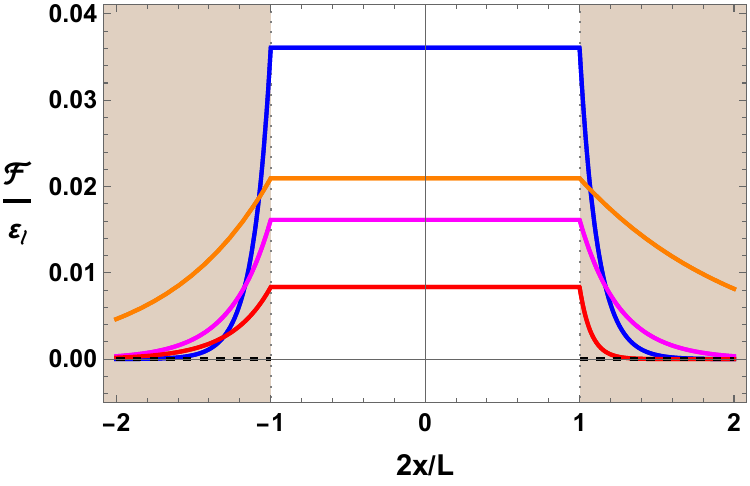}
\caption{Stationary profiles of the energy density (left) and the energy flux (right) for an ideal ultrarelativistic fluid without a conserved charge confined between two thermal walls (shaded regions), which absorb and emit particles according to the Kirchhoff-Planck law~\eqref{wallcollisions}. We take $\varepsilon_{\mathit r}=7\varepsilon_\ell/8$ (so that $|\mathcal{F}|\ll \varepsilon$, which is needed for \eqref{idealfluid} to hold). The solid curves show the exact stationary solution~\eqref{ViscStatSolut} for $\{D/L,\tau_\ell/L,\tau_{\mathit r}/L\}=\{0,0.1,0.1\}$ (blue), $\{0.1,0.1,0.1\}$ (magenta), $\{0.1,0.1,0.01\}$ (red), and $\{0.2,0.3,0.6\}$ (orange), while the dashed lines indicate the corresponding black-body values deep inside each wall. The presence of viscosity allows for a discontinuity in the energy density at the interfaces, while preserving continuity of the total normal stress. The value of $\varepsilon$ in the gap is not the mean of the wall black bodies, but is shifted toward that of the wall with the larger absorption length, and the stationary energy flux is reduced.
}
\label{fig:StationaryVisc}
\end{figure}

We seek stationary solutions $\{\varepsilon_0(x),\mathcal{F}_0(x)\}$ that approach the corresponding black-body equilibrium states as $x\to\pm\infty$. The solution must also satisfy appropriate matching conditions at the interfaces $x=\pm L/2$. Integrating Eqs.~\eqref{ViscosoneLong} across an infinitesimal neighborhood of each interface shows that $\mathcal{F}$ is continuous, while the combination $\frac{1}{3}\varepsilon-\frac{4}{3}D\,\partial_x\mathcal{F}$
must also be continuous across the interfaces. As a result, the energy density itself need not be continuous: any jump in $\varepsilon$ is exactly compensated by an equal and opposite jump in the viscous stress, ensuring that the total normal stress remains continuous across the interface. Naturally, these conditions simply reflect that both the energy flux $T^{0x}$ and the normal stress $T^{xx}$ are continuous across the interfaces.
Solving Eqs.~\eqref{ViscosoneLong} subject to these conditions yields
(see figure \ref{fig:StationaryVisc}), 
\vspace{-0.1cm}
\begin{equation}\label{ViscStatSolut}
\begin{split}
\varepsilon_0={}& a
\begin{cases}
T_\ell^4 +\dfrac{T_{\mathit r}^4-T_{\ell}^4}{\sqrt{1{+}\frac{4D}{\tau_\ell}} \left(\sqrt{1{+}\frac{4D}{\tau_\ell}}+\sqrt{1{+}\frac{4D}{\tau_\mathit{r}}}\right)} \exp{\left[\dfrac{\sqrt{3}\, (x{+}\frac{L}{2})}{\tau_\ell \sqrt{1{+}\frac{4D}{\tau_\ell}}}\right]}, & x\,{\le}\, {-}L/2\,,\\[20pt]
\dfrac{T_{\ell}^4\sqrt{1{+}\frac{4D}{\tau_\mathit{r}}}+T_{\mathit r}^4 \sqrt{1{+}\frac{4D}{\tau_\ell}}}{\sqrt{1{+}\frac{4D}{\tau_\ell}}+\sqrt{1{+}\frac{4D}{\tau_\mathit{r}}}}, & {-}L/2\,{<}\,x\,{<}\,L/2\,,\\[10pt]
T_\mathit{r}^4 +\dfrac{T_{\ell}^4-T_{\mathit r}^4}{\sqrt{1{+}\frac{4D}{\tau_\mathit{r}}} \left(\sqrt{1{+}\frac{4D}{\tau_\ell}}+\sqrt{1{+}\frac{4D}{\tau_\mathit{r}}}\right)}\exp{\left[-\dfrac{\sqrt{3}\, (x{-}\frac{L}{2})}{\tau_{\mathit r} \sqrt{1{+}\frac{4D}{\tau_\mathit{r}}}}\right]}, & x\ge L/2\, ,
\end{cases}\\
\mathcal{F}_0={}& \dfrac{a(T_\ell^4{-}T_\mathit{r}^4)}{\sqrt{3}\left(\sqrt{1{+}\frac{4D}{\tau_\ell}}+\sqrt{1{+}\frac{4D}{\tau_\mathit{r}}}\right)}
\begin{cases}
\exp{\left[\dfrac{\sqrt{3}\, (x{+}\frac{L}{2})}{\tau_\ell \sqrt{1{+}\frac{4D}{\tau_\ell}}}\right]}, & x\le -L/2\,,\\[4pt]
1, & -L/2<x<L/2\,,\\[4pt]
\exp{\left[-\dfrac{\sqrt{3}\, (x{-}\frac{L}{2})}{\tau_{\mathit r} \sqrt{1{+}\frac{4D}{\tau_\mathit{r}}}}\right]}, & x\ge L/2\, .
\end{cases}\\
\end{split}
\end{equation}

The stationary solution exhibits several interesting features. Between the plates, both the energy density and the energy flux are spatially uniform. In contrast to the non-self-interacting case \eqref{emomentumstefan}, however, the energy density does not coincide with the arithmetic mean of the two black-body energy densities. Instead, it is shifted toward the equilibrium value of the wall with the larger absorption length, reflecting the stronger influence of that reservoir on the stationary state. Within the walls, the energy density and energy flux both relax exponentially toward their respective equilibrium values. The presence of viscosity modifies this relaxation by increasing the characteristic penetration length into the walls, thereby broadening the boundary layers over which the fluid equilibrates with the reservoirs. At the same time, the stationary energy flux decreases as the diffusivity increases, reflecting the additional resistance to energy transport introduced by viscous momentum diffusion.

\begin{figure}[h!]
    \centering
\includegraphics[width=0.44\linewidth]{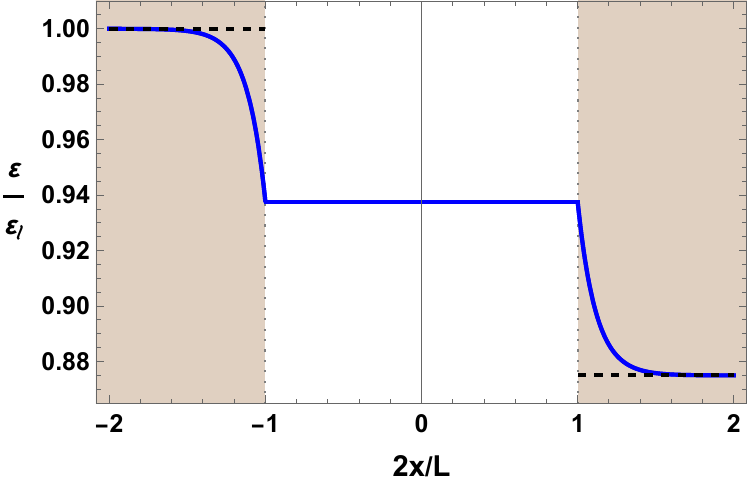}\hspace{0.05\linewidth}
\includegraphics[width=0.44\linewidth]{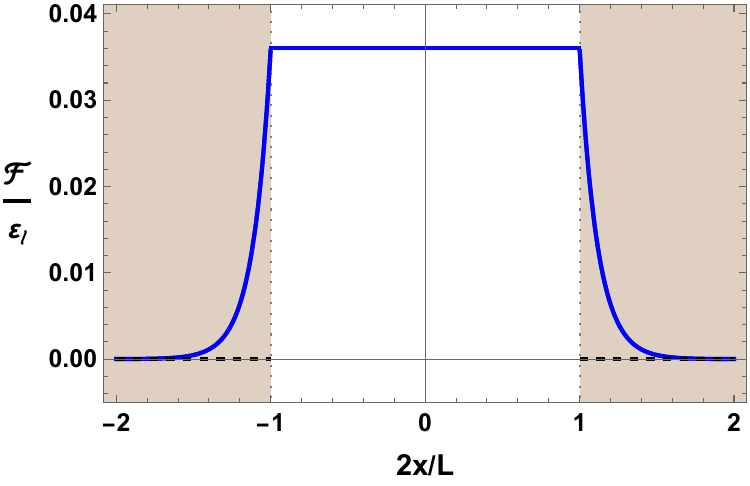}
\caption{Same as in Fig. \ref{fig:StationaryVisc}, but with no viscosity ($D=0$). The solid blue curves show the exact stationary solution~\eqref{InvishStatSolut}, while the dashed lines indicate the corresponding black-body values deep inside each wall. The fluid temperature and flow velocity do not match those of the walls at the interfaces; instead, equilibration occurs within boundary layers inside the walls, with characteristic thickness set by the absorption length $\tau_{\ell/\mathit r}$.
}
\label{fig:StationaryInviscid}
\end{figure}

An ideal-fluid solution is recovered by taking the limit $D\rightarrow0$ in \eqref{ViscStatSolut} (see figure \ref{fig:StationaryInviscid}):
\begin{equation}\label{InvishStatSolut}
\varepsilon_0=a
\begin{cases}
T_\ell^4 +\frac{T_{\mathit r}^4-T_{\ell}^4}{2} e^{\frac{\sqrt{3}}{\tau_\ell}(x+\frac{L}{2})}, & x\,{\le}\, {-}L/2\,,\\[4pt]
\frac{T_{\ell}^4+T_{\mathit r}^4}{2}, & {-}L/2\,{<}\,x\,{<}\,L/2\,,\\[4pt]
T_\mathit{r}^4 +\frac{T_{\ell}^4-T_{\mathit r}^4}{2} e^{-\frac{\sqrt{3}}{\tau_{\mathit r}}(x-\frac{L}{2})}, & x\,{\ge}\, L/2\, ,
\end{cases}
\qquad
\mathcal{F}_0= \dfrac{a(T_\ell^4{-}T_\mathit{r}^4)}{2\sqrt{3}}
\begin{cases}
e^{\frac{\sqrt{3}}{\tau_\ell}(x+\frac{L}{2})}, & x\le -L/2\,,\\[4pt]
1, & -L/2<x<L/2\,,\\[4pt]
e^{-\frac{\sqrt{3}}{\tau_{\mathit r}}(x-\frac{L}{2})}, & x\ge L/2\, .
\end{cases}
\end{equation}
In this limit, the energy density in the gap reduces to the arithmetic mean of the two black-body energy densities, while the penetration length inside each wall becomes $\tau_{\ell,\mathit r}/\sqrt{3}$. The appearance of the factor $\sqrt{3}$ in the relaxation length reflects the angular structure of an isotropic ultrarelativistic distribution (for a perfectly collimated beam propagating along $x$, the relaxation length would reduce to $\tau_{\ell/\mathit r}$). Moreover, the stationary energy flux attains its maximum value, since momentum is no longer dissipated through viscous diffusion. Interestingly, even in the absence of viscosity, the stationary energy flux does not coincide with the Stefan–Boltzmann value. Instead, it changes by a factor $2/\sqrt{3}$.

Note that, in the ideal-fluid limit, the energy density is continuous at the interface.

\newpage
At first sight, the existence of a stationary energy current in the ideal-fluid limit appears paradoxical, since heat transport is commonly associated with a dissipative process (diffusion). This intuition is rooted in the Eckart description, where heat is represented by a diffusive current relative to the particle flow. In the present problem, however, there is no conserved particle current and the Landau frame is the most natural hydrodynamic frame. Energy transport is therefore encoded directly in the fluid velocity itself. The exchange of particles with the reservoirs generates a stationary flow that carries energy from one wall to the other, while viscosity merely modifies this flow through the momentum balance. The limit $D\rightarrow0$ is thus not the disappearance of energy transport, but the removal of viscous resistance to a boundary-driven stationary flow.

\vspace{-0.6cm}
\subsection{Relaxation of fluctuations}
\vspace{-0.3cm}

\begin{figure}[h!]
    \centering
\includegraphics[width=0.44\linewidth]{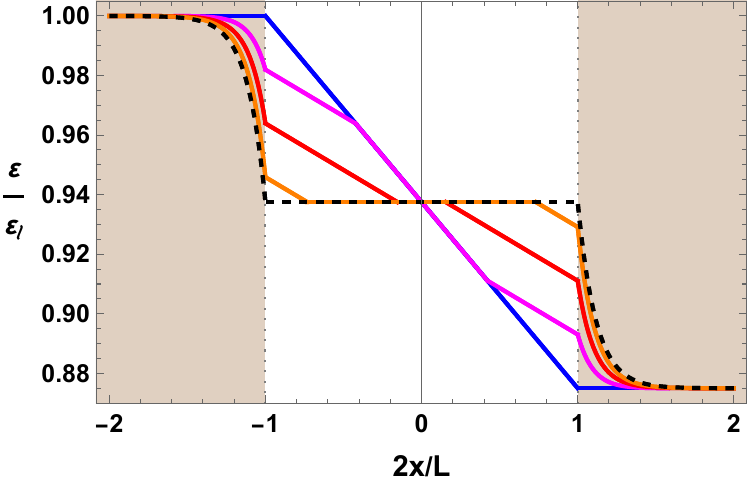}\hspace{0.05\linewidth}
\includegraphics[width=0.44\linewidth]{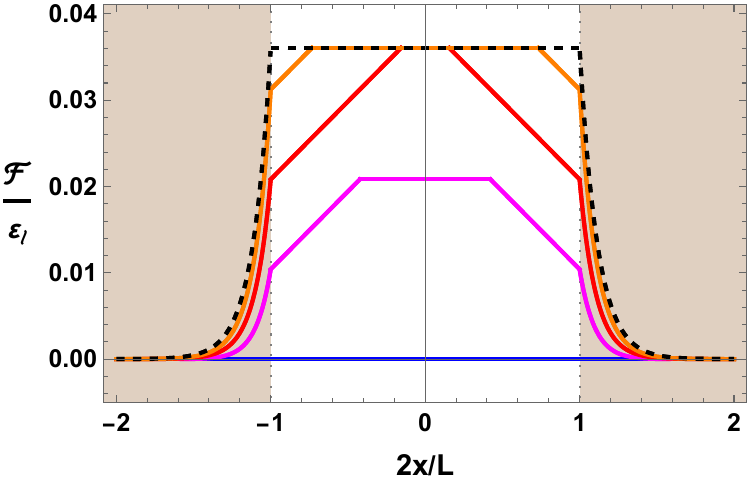}
\includegraphics[width=0.44\linewidth]{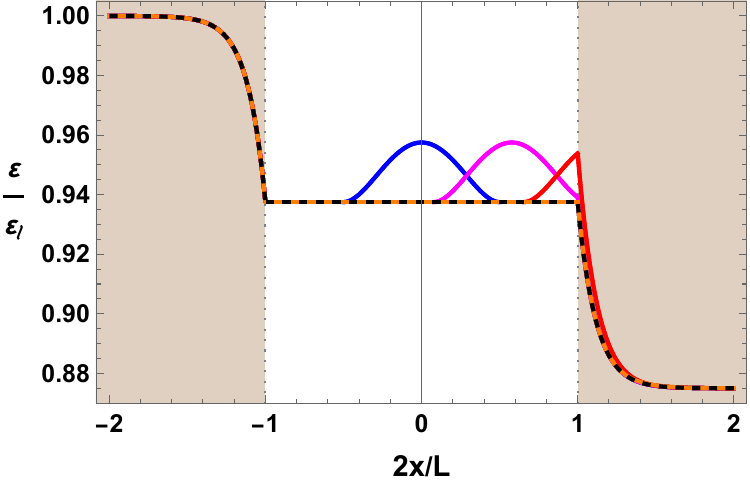}\hspace{0.05\linewidth}
\includegraphics[width=0.44\linewidth]{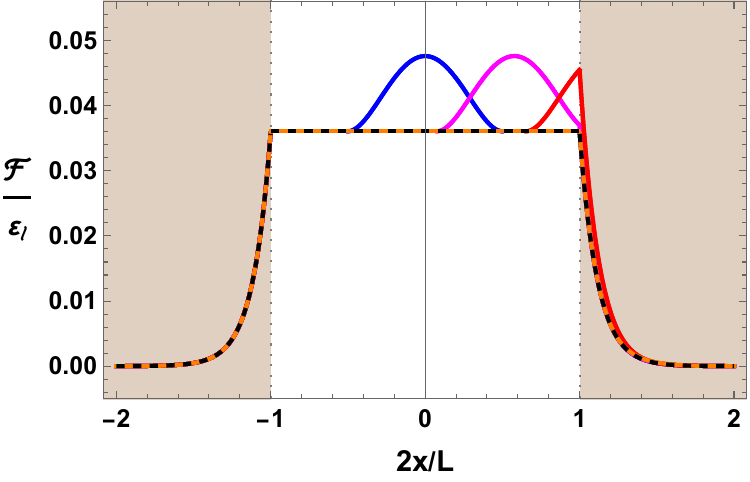}
\caption{Evolution of the energy density (left panels) and energy flux (right panels) for a fluid under the same external conditions as in Fig.~\ref{fig:StationaryInviscid}, but for different initial data. The solid curves correspond to snapshots of the exact solution~\eqref{planarfluctuation} at different times (blue: $2t/L{=}0$, magenta: $2t/L{=}1$, red: $2t/L{=}2$, orange: $2t/L{=}3$), while the dashed curve indicates the stationary solution~\eqref{InvishStatSolut}. 
Upper panels: the fluid is initially at rest with a prescribed energy ramp, similar to standard Newtonian setups. The resulting pressure gradient drives an accelerating flow that relaxes toward the stationary profile. 
Lower panels: the fluid is initialized with a right-moving perturbation, which propagates across the gap and is absorbed by the wall without reflection.
}
    \label{fig:RampandBump}
\end{figure}

Let us now focus on the fluctuation $\{\Delta \varepsilon,\Delta \mathcal{F}\}$, which solves the homogeneous version of equation \eqref{ViscosoneLong} and decays at infinity. For the sake of simplicity, we consider these solutions in the ideal fluid limit. Using the method of the characteristics, we obtain the following general solution (see figure \ref{fig:RampandBump}):
\vspace{-0.1cm}
\begin{equation}\label{planarfluctuation}
\begin{pmatrix}
\Delta \varepsilon \\
\Delta \mathcal{F}  \\
\end{pmatrix}
=\psi_\mathit{r}(x{-}t/\sqrt{3})\begin{pmatrix}
1\\
1/\sqrt{3}\\
\end{pmatrix} \times \begin{cases}
e^{-\frac{\sqrt{3}}{\tau_\ell}(x+\frac{L}{2})}\\[1pt]
1\\[1pt]
e^{-\frac{\sqrt{3}}{\tau_{\mathit r}}(x-\frac{L}{2})}
\end{cases}
+\psi_\ell (x{+}t/\sqrt{3}) \begin{pmatrix}
1\\
-1/\sqrt{3}\\
\end{pmatrix} \times \begin{cases}
e^{\frac{\sqrt{3}}{\tau_\ell}(x+\frac{L}{2})}, & x\le -L/2\,,\\[1pt]
1, & -L/2<x<L/2\,,\\[1pt]
e^{\frac{\sqrt{3}}{\tau_{\mathit r}}(x-\frac{L}{2})}, & x\ge L/2\, .
\end{cases}
\end{equation}
This solution is a superposition of a right-moving wave $\psi_{\mathit r}$ and a left-moving wave $\psi_{\ell}$, both propagating at the sound speed $1/\sqrt{3}$. In the region between the walls, the amplitudes remain constant. Inside the walls, however, the modes are exponentially attenuated, reflecting the damping induced by absorption. No reflection occurs at the interfaces: the functions $\psi_{\ell}$ and $\psi_{\mathit r}$ are freely specifiable and evolve independently.

\subsection{Effective boundary-value problem}
\vspace{-0.3cm}

We now derive the effective boundary-value problem governing the evolution of perturbations inside the gap between the walls. Although the microscopic dynamics is defined over the entire domain, including the wall regions, the exponential attenuation of perturbations inside the walls allows this description to be reduced to an equivalent problem posed solely on the interval $-L/2<x<L/2$.

Suppose that the initial fluctuation $\{\Delta\varepsilon,\Delta\mathcal{F}\}$ is supported entirely within the gap at $t=0$. Then, the characteristic amplitudes $\psi_{\mathit r}$ and $\psi_\ell$ are likewise initially supported in this region. Since the right-moving mode propagates exclusively toward increasing $x$, it never reaches the left boundary. Consequently, the perturbation at $x=-L/2$ consists entirely of the left-moving mode,
\begin{equation}
\{\Delta\varepsilon,\Delta\mathcal{F}\}
=
\psi_\ell
\left\{
1,-\frac{1}{\sqrt{3}}
\right\}.
\end{equation}
Similarly, the left-moving mode propagates exclusively toward decreasing $x$ and never reaches the right boundary, so that the perturbation at $x=L/2$ is entirely given by the right-moving mode,
\begin{equation}
\{\Delta\varepsilon,\Delta\mathcal{F}\}
=
\psi_{\mathit r}
\left\{
1,\frac{1}{\sqrt{3}}
\right\}.
\end{equation}
Therefore, when restricting attention to perturbations initially localized inside the gap, the evolution is completely described by the effective boundary conditions
\begin{equation}
\left(\Delta \mathcal{F}+\frac{\Delta\varepsilon}{\sqrt{3}}\right)_{x=-L/2}=0\,, \qquad \qquad
\left(\Delta \mathcal{F}-\frac{\Delta\varepsilon}{\sqrt{3}}\right)_{x=+L/2}=0\, .
\label{maximallyabsorbing}
\end{equation}
These conditions simply require the incoming characteristic fields to vanish at the boundaries. In the theory of hyperbolic systems, they are known as \textit{maximally absorbing boundary conditions}, since they eliminate incoming waves while allowing outgoing perturbations to leave the domain without reflection. Such boundary conditions are well known to yield a well-posed initial-boundary-value problem \cite{Sarbach:2012pr}.

Existence follows directly by restricting the explicit solution~\eqref{planarfluctuation} to the interval between the walls. To establish uniqueness and continuous dependence on the initial data, consider the quadratic functional
\begin{equation}
E(t)=\frac{1}{2}\int_{-L/2}^{L/2}
\left[
(\Delta\varepsilon)^2
+
3(\Delta\mathcal{F})^2
\right]
dx ,
\label{infona}
\end{equation}
which coincides with the perturbation of the Helmholtz free energy contained within the gap. Differentiating $E(t)$ with respect to time and using \eqref{ViscosoneLong} (with $D=0$) together with \eqref{maximallyabsorbing} yields $\dot E
=
-\frac{1}{\sqrt{3}}
(\Delta\varepsilon)^2_{x=-L/2}
-
\frac{1}{\sqrt{3}}
(\Delta\varepsilon)^2_{x=+L/2}
\le0.$
Hence, $E(0)=0$ implies $E(t)=0$, establishing uniqueness, while $E(t)\le E(0)$ guarantees continuous dependence on the initial data.

In summary, the microscopic interaction between the fluid and the thermally conducting walls reduces, at the hydrodynamic level, to an effective boundary-value problem posed entirely within the gap. The resulting boundary conditions are maximally absorbing, reflecting the fact that perturbations propagating into the walls decay exponentially and do not generate incoming hydrodynamic modes. Consequently, the associated initial-boundary-value problem is well posed. Thus, the microscopic wall model not only determines the stationary state, but also uniquely fixes the physically appropriate boundary conditions governing the evolution of longitudinal perturbations.

\vspace{-0.3cm}
\subsection{Comparison to numerical solution of the Boltzmann equation
}
\vspace{-0.3cm}

Next, we will investigate how numerical solutions to the relativistic Boltzmann equation behave in comparison to the ideal hydrodynamic and free-streaming limits. We consider the interactions between bosons to be described in the Anderson-Witting relaxation time approximation (RTA)~\cite{AndersonWitting1974}, such that the collision integral $\mathcal{C}_{bb}$ is given by
\begin{align}\label{rtaCbb}
    \mathcal{C}_{bb}=\frac{p \cdot u(x)}{\tau_R(x)} \left[ f(x, p) - f_{\rm eq}(T(x),u_{\mu}p^{\mu}) \right]
\end{align}
where in contrast to the boson-wall interaction (see \cref{wallcollisions}), the equilibrium distribution on the right-hand side of \cref{rtaCbb} is self-consistently determined by the local temperature $T(x)$ and velocity  $u^{\mu}(x)$ of the fluid via the Landau matching condition
\begin{align}
    T^{\mu\nu}(x)u_{\nu}(x)=-e(T(x))u^{\mu}(x)
\end{align}
We consider a conformal system, with equation of state $\varepsilon=3P=\frac{\pi^2}{30}T^{4}$ characteristic of massless bosons (i.e. $a=\frac{\pi^2}{30}$) and constant shear-viscosity to entropy density $\eta/s$, such that the relaxation time of the fluid is determined by the local temperature $T$ of the fluid, according to $\tau_R=5 (\eta/s) T^{-1}$. Details of the numerical implementation are provided in Appendix \ref{ccccccccc}, and the software is available as part of this submission.

\begin{figure}[h!]
    \centering
\includegraphics[width=0.32\linewidth]{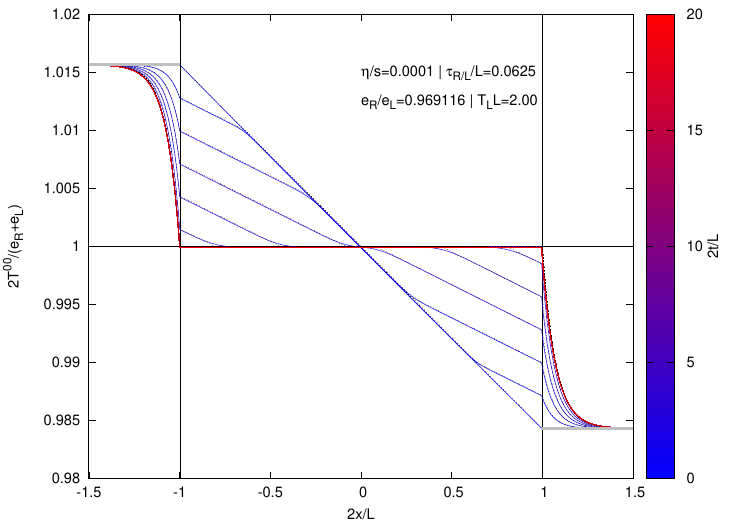}
\includegraphics[width=0.32\linewidth]{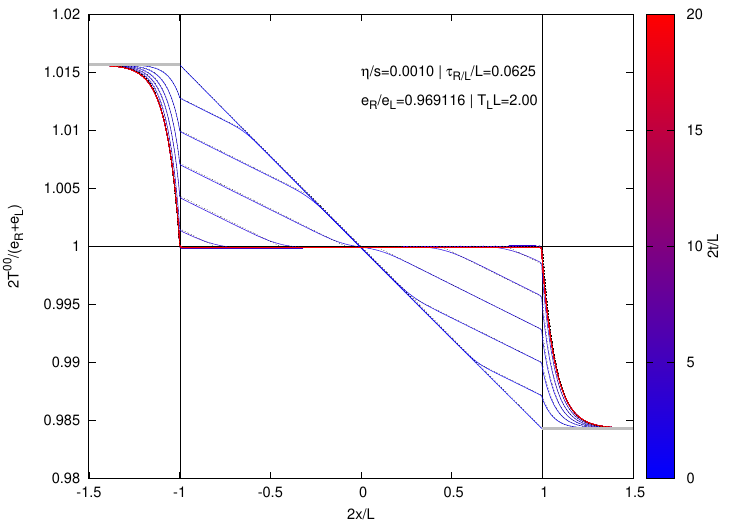}
\includegraphics[width=0.32\linewidth]{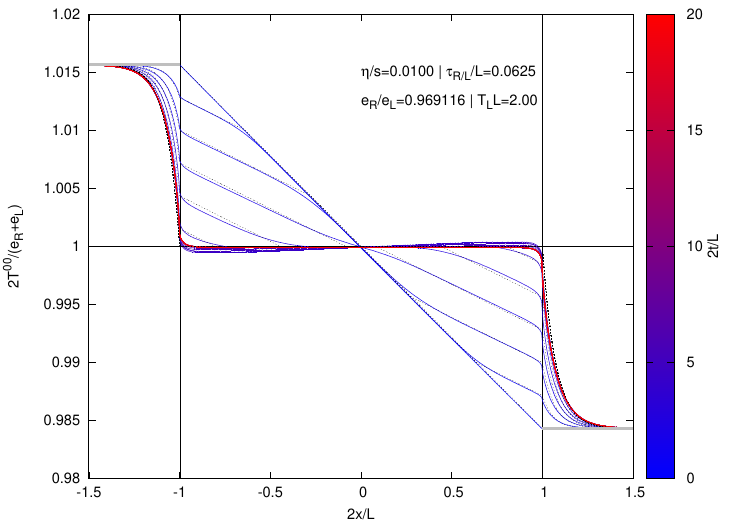}
\includegraphics[width=0.32\linewidth]{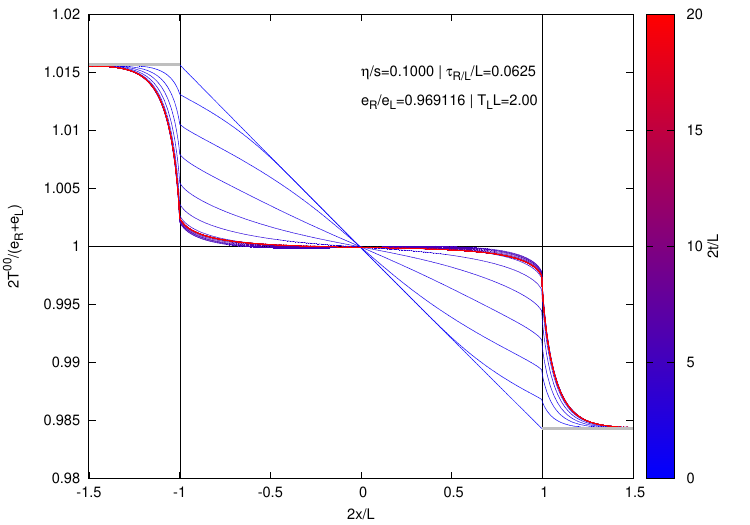}
\includegraphics[width=0.32\linewidth]{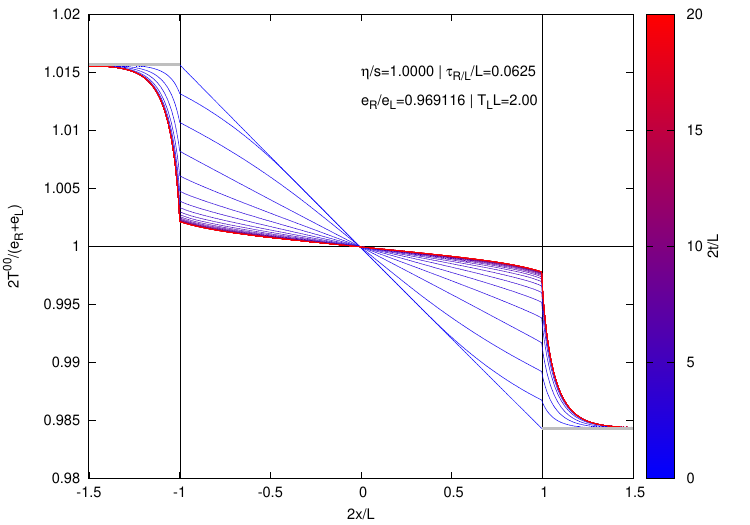}
\includegraphics[width=0.32\linewidth]{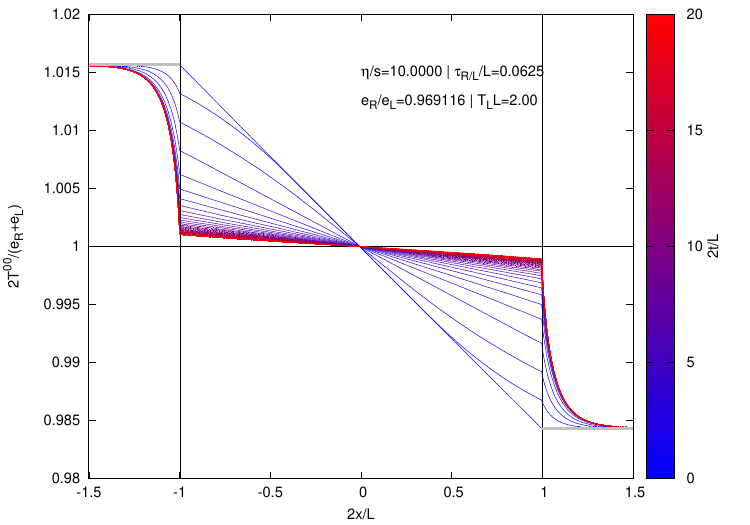}
\includegraphics[width=0.32\linewidth]{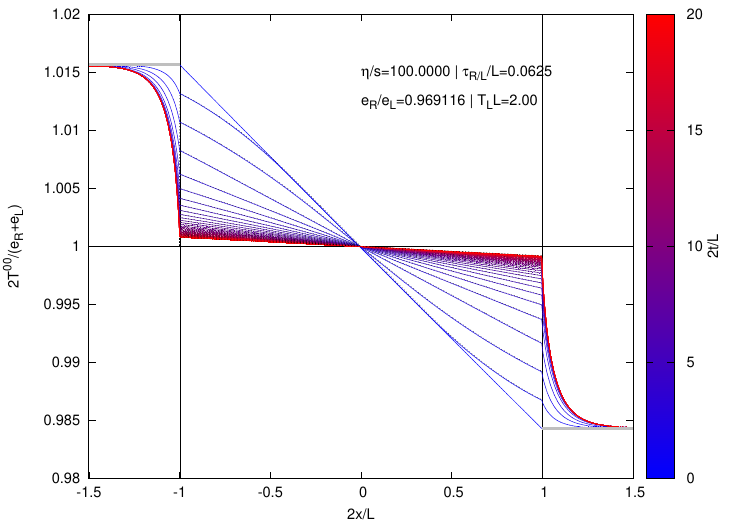}
\includegraphics[width=0.32\linewidth]{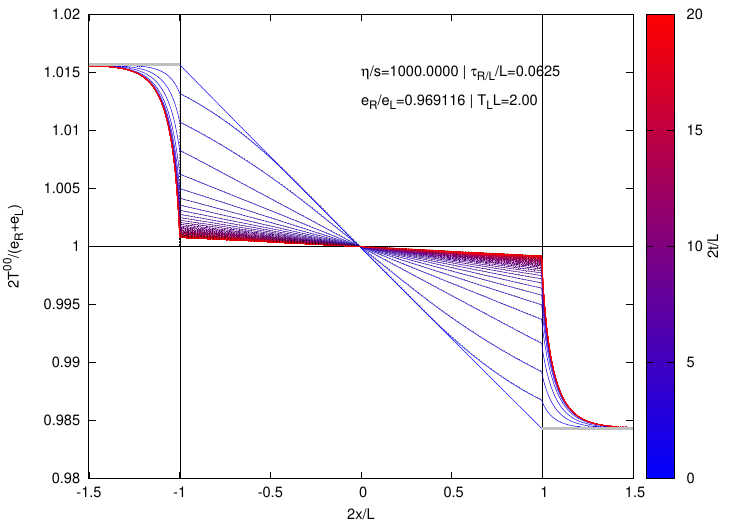}
\includegraphics[width=0.32\linewidth]{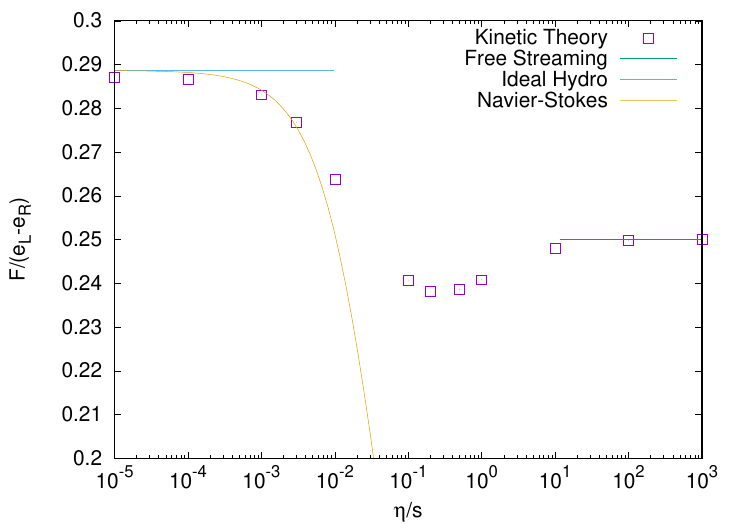}

\caption{Evolution of the energy density in kinetic theory. Different panels show the result across a large range of viscosities $\eta/s=10^{-4} - 10^{3}$. Dashed curves in the top left and bottom panels show the comparison to the ideal hydro and free-streaming solutions. (bottom right) Stationary energy flux $\mathcal{F}$  from the hot to the cold reservoir, normalized by the difference of the equilibrium energy density $e_l-e_r=a(T_{l}^4-T_{r}^4)$ as a function of viscosity $\eta/s$, compared to the free-streaming limit (c.f. \cref{emomentumstefan}), as well as the ideal and viscous hydrodynamic solutions (c.f. \cref{InvishStatSolut,ViscStatSolut}).}
\label{fig:Numerics1}. 
\end{figure}

We provide a compact summary of our results in Fig.~\ref{fig:Numerics1}, where we present the time evolution of the solution for $T^{00}$ across a large range of viscosities $\eta/s=10^{-4} \to 10^{3}$ (top left to bottom right panels). Starting from very small viscosities $\eta/s \ll 1$, where the numerical solution shows the ideal hydrodynamic behavior, one observes how a larger viscosity of the fluid smooths the energy density profile close to the boundaries. Eventually, when viscosity increases further $\eta/s \gtrsim 1$, the profile becomes sharper again, as the solution converges towards the free-streaming limit (see Appendix \ref{freestreaimng} for the analytical free-streaming solution). Besides the evolution of the profiles, the bottom right panel of Fig.~\ref{fig:Numerics1} also shows a comparison of the (stationary) energy flux $\mathcal{F}$ from the hot to the cold reservoir, obtained at a late time $t/L=100$, where the transient dynamics has converged towards the stationary solution. We find that the energy flux in kinetic theory smoothly interpolates between the ideal hydrodynamic and free-streaming limit, and attains a minimum around $\eta/s \sim 1$, when the internal relaxation time $\tau_R$ of the fluid becomes comparable to the distance $L$ between the two plates.

% While the stationary solutions for the $T^{00}$ profile agree in the ideal hydrodynamic and free-streaming limits, one naturally expects viscosity to affect the profiles at intermediate viscosities, as can be seen from Fig. \ref{fig:Numerics2} where we present a comparison of the stationary profiles of $T^{00}$, obtained numerically by following the evolution for a very long time ($t/L=100$), for small and large viscosities in the left and central panels.

\begin{figure}[h!]
    \centering
\includegraphics[width=0.45\linewidth]{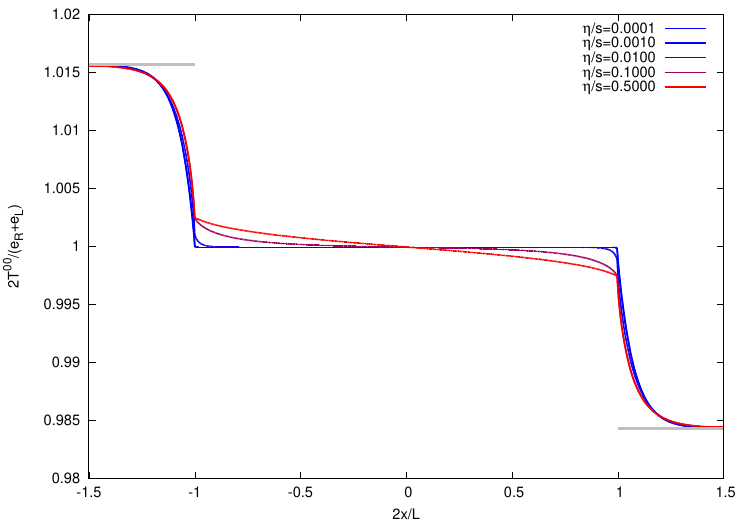}
\includegraphics[width=0.45\linewidth]{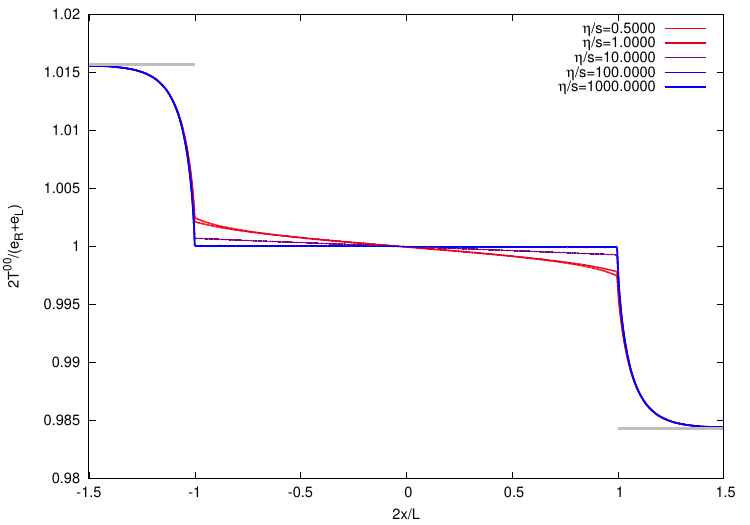}
\caption{Stationary energy density profiles obtained from numerical solution of the relativistic Boltzmann equation for different viscosities $\eta/s=10^{-4} - 0.5$ (left) and $0.5 - 10^{3}$ (right). }
\label{fig:Numerics2}
\end{figure}

\subsection{Failure of the Navier--Stokes approximation near the walls}
We can also directly compare the stationary profiles of $T^{00}$, which we obtain numerically by following the evolution for a very long time ($t/L=100$), as depicted in  Fig. \ref{fig:Numerics2} for small and large viscosities in the left and right panels. We observe from \cref{fig:Numerics2} that,
for small shear viscosity, $\eta/s \ll 1$, the Boltzmann equation converges to the ideal-fluid profile \eqref{InvishStatSolut}, reproducing both the uniform bulk value of $T^{00}$ and its non-differentiable behavior at the interfaces (see also top-left panel of Fig.~\ref{fig:Numerics1}). Here ideal hydrodynamics performs remarkably well. 

However, the situation changes as soon as dissipative corrections become non-negligible for larger viscosities. Here the comparison between the kinetic theory and viscous hydrodynamics exposes a qualitative limitation of the Navier-Stokes approximation. The Navier-Stokes solution \eqref{ViscStatSolut} \emph{sharpens} the boundary layer as $D$ (equivalently $\eta/s$) increases, rather than smoothing it: the energy density becomes discontinuous across each interface, the jump being compensated by an equal and opposite jump in $\pi^{xx}$ so that only the total normal stress $T^{xx}$ stays continuous. The kinetic solution behaves oppositely: as $\eta/s$ grows, the Boltzmann $T^{00}$ profile never develops a discontinuity, but instead smooths out over a boundary layer of finite width (Figs.~\ref{fig:Numerics1}--\ref{fig:Numerics2}). Thus, precisely where dissipative corrections matter most, the viscous hydrodynamic solution is qualitatively \emph{less regular} than the microscopic solution it approximates (a sign that the Navier-Stokes truncation is not just quantitatively off near the walls, but structurally the wrong kind of solution).

The reason is not hard to identify. The Navier-Stokes solution \eqref{ViscStatSolut} solves the hydrodynamic equations with the wall's absorption/emission term treated as a fixed, spatially discontinuous source: it switches on abruptly at $x=\pm L/2$, jumping from zero in the gap to $O(\mathcal{F}/\tau_{\ell,r})$ inside the wall, regardless of how large or small $\tau_{\ell,r}$ is. Since a first-order gradient expansion inherits the regularity of its source, the resulting Navier-Stokes profile for $\varepsilon$ is itself discontinuous. In contrast, the kinetic solution has no such feature, because the microscopic process is not an instantaneous local switch: particles absorbed and re-emitted by the wall retain memory of the angular distribution with which they arrived, which smooths the transition. The Navier-Stokes constitutive relation \eqref{NavierStokesSress}, being purely local and sensitive only to the instantaneous source, cannot encode this memory. Truncating the gradient expansion at first order therefore does not give a small correction to the ideal-fluid result near the interface; it enforces a discontinuity exactly where the true solution is smooth (the signature of a gradient expansion applied to a source for which no such expansion holds).

This matches the general picture of boundary-layer solutions in relativistic kinetic theory from Ref.~\cite{GavassinoInterfaces:2026fkj}, where disturbances localized near a planar interface moving at fixed velocity are built from modes with purely imaginary frequency and wavenumber, selected by matching to the interface's worldline in the $(i\omega,ik)$ plane (see also \cite{GavassinoLorenztian:2026seq}). At low wall velocities these modes lie off the hydrodynamic branches and cannot be captured by any finite truncation of the gradient expansion. In this language, the tails of our Boltzmann solution near $x=\pm L/2$ are purely non-hydrodynamic. 
They do not represent small deviations from local equilibrium, but rather correspond to boundary-layer excitations sourced by the wall itself, which no local constitutive relation (Navier-Stokes, Israel-Stewart, or otherwise) can reproduce once the boundary layer is resolved on hydrodynamic length scales. Concretely, these tails are a continuous superposition of exponentials as determined by the spectral decomposition of the propagator in \cite{GavassinoInterfaces:2026fkj}, which cannot be captured by the finite number of discrete modes that hydrodynamics inevitably relies on.

This is a statement about the derivative expansion \emph{near the boundary}, not about the bulk stationary flow, which remains well described by ideal hydrodynamics away from the walls. It suggests, however, that a faithful hydrodynamic treatment of thermally conducting boundaries (beyond the ideal-fluid level) cannot be built by supplementing the bulk equations with a local Navier-Stokes closure up to the wall; an explicitly kinetic treatment of the boundary layer appears unavoidable. A formal analytical treatment along these lines will be carried out in a follow-up paper.

\vspace{-0.3cm}
\section{Conclusions}
\vspace{-0.3cm}

In this work, we investigated the interaction between an ultra-relativistic fluid without a conserved particle number and thermally conducting walls. Rather than prescribing boundary conditions directly at the hydrodynamic level, we modeled the reservoirs microscopically as black-body emitters and absorbers of massless bosons. This framework provides a physically motivated description of energy exchange between the fluid and its surroundings and allows the corresponding hydrodynamic boundary conditions to be derived rather than postulated.

We first analyzed the stationary solutions of the resulting kinetic model. In contrast to the conventional picture based on fixed wall temperatures, we found that the stationary state is characterized by a nonvanishing particle flux between the reservoirs, sustained by continuous particle emission and absorption at the walls. This mechanism maintains an approximately uniform pressure throughout the system while transporting energy from the hotter reservoir to the colder one. Viscosity modifies the quantitative properties of the stationary state, increasing the penetration depth of the boundary layers and reducing the energy flux, while smoothly recovering the inviscid limit.

We then investigated the propagation of longitudinal perturbations about these stationary configurations. Solving the linearized equations over the entire domain, including the wall regions, showed that perturbations entering the walls decay exponentially due to particle absorption. As a consequence, the dynamics within the fluid is completely described by an effective initial-boundary-value problem posed only inside the gap. The corresponding boundary conditions are precisely the maximally absorbing conditions familiar from the theory of hyperbolic systems, implying that the effective problem is well posed. In this way, the microscopic interaction between the fluid and the reservoirs uniquely determines the appropriate hydrodynamic boundary conditions governing the evolution of perturbations.

Finally, we compared the hydrodynamic predictions to direct numerical solutions of the Boltzmann equation across a wide range of viscosities. While the ideal-fluid limit is reproduced essentially exactly, the first-order Navier-Stokes approximation fails qualitatively near the walls: it enforces a discontinuous energy density at the interfaces, whereas the true kinetic solution is smooth there. We traced this failure to the discontinuous nature of the wall's absorption/emission source term, which a local, first-order gradient expansion cannot resolve, since the true kinetic profile retains memory of the angular distribution of particles absorbed and re-emitted by the wall. This indicates that the boundary layers are intrinsically non-hydrodynamic, in the sense that they cannot be captured by any local constitutive relation, however many gradients are included, and that their proper description requires an explicit kinetic (or matched-asymptotic) treatment.

The physical picture that emerges differs qualitatively from the standard description of heat transport in relativistic hydrodynamics. In the absence of a conserved particle current there is no Eckart frame and the Landau frame therefore provides the unique hydrodynamic description. Stationary energy transport is therefore necessarily accompanied by a nonvanishing fluid velocity. From this perspective, the particle flow established between the reservoirs is not an artifact of the model, but rather the natural hydrodynamic manifestation of heat transport in a fluid whose microscopic constituents are continuously exchanged with the walls.

Beyond the specific model considered here, our results suggest a more general perspective on boundary conditions in relativistic hydrodynamics. Rather than prescribing boundary conditions directly for the hydrodynamic variables, one may formulate the interaction with the environment microscopically and derive the corresponding effective hydrodynamic description. While the present work focused on fluids at vanishing chemical potential, the same philosophy should extend to systems with conserved charges, where physically motivated boundary conditions in the Landau frame remain largely unexplored. We hope that the framework developed here provides a useful starting point for such investigations.

\vspace{-0.3cm}
\section*{Acknowledgements}
\vspace{-0.3cm}

LG is supported by a MERAC Foundation prize grant,  an Isaac Newton Trust Grant, and funding from the Cambridge Centre for Theoretical Cosmology. S.S. was supported by the
Deutsche Forschungsgemeinschaft (DFG, German Research Foundation) through the CRC-TR 211 ‘Strong interaction matter under extreme conditions’-project number 315477589 – TRR 211. G.S.D.~is supported by CNPq through the grant 307761/2022-3 and acknowledges the support of the INCT-FNA grant 408419/2024-5. The authors gratefully acknowledge the computing time made available to them on the high-performance computer Otus at the NHR Center Paderborn Center for Parallel Computing (PC2). This center is jointly supported by the Federal Ministry of Research, Technology and Space and the state governments participating in the National High-Performance Computing (NHR) joint funding program (www.nhr-verein.de/en/our-partners).

\appendix

\section{Dirichlet solution with unequal temperatures}\label{aaa}

\subsubsection{Without viscosity}
Let us consider a zero-chemical-potential fluid with $\varepsilon=3P$ (and hence $P\propto T^4$), and with vanishing viscosity. Then, the stress-energy tensor takes the ideal-fluid form
\begin{equation}
T^{\mu \nu}=3P u^\mu u^\nu +P \Delta^{\mu \nu}\, ,
\end{equation}
with $\Delta^{\mu \nu}=\eta^{\mu \nu}+u^\mu u^\nu$. Suppose that this fluid is placed between two plates located at $x=\pm L/2$ and having temperature $T_\pm$, and let us look for stationary planar-symmetric solutions satisfying the Dirichlet data $T(\pm L/2)=T_{\pm}$. Then, taking the flow velocity to be $u^\mu(x)=(\gamma,u_x,0,0)$, the conservation law $\partial_\mu T^{\mu \nu}=0$ yields
\begin{equation}\label{ABIdeal}
\begin{split}
\partial_x T^{xt}=0 & \qquad \Longrightarrow \qquad 4P \gamma u_x=A \, , \\
\partial_x T^{xx}=0 & \qquad \Longrightarrow \qquad 4P u_x^2+P=B \, . \\
\end{split}
\end{equation}
Since these equations are algebraic, there is no continuous solution that connects two different boundary temperatures $T_-$ and $T_+$. Consequently, the fluid can satisfy the boundary conditions $T(\pm L/2)=T_\pm$ only by forming a shock. Indeed, the two relations in \eqref{ABIdeal} are just the Rankine-Hugoniot conditions \cite[Eq.~(2.7.2)]{NovikovThorne1973}. To determine the solution for given boundary data, one needs to arbitrarily specify the location $x_0\in (-L/2,L/2)$ of the discontinuity, and set
\begin{equation}\label{discontinuoussolution}
P(x)=
\begin{cases}
P_- & \text{for }x<x_0 \, ,\\
P_+ & \text{for }x>x_0 \, , \\
\end{cases}\qquad
u_x(x)=
\begin{cases}
u_- & \text{for }x<x_0 \, , \\
u_+ & \text{for }x>x_0 \, , \\
\end{cases}
\end{equation}
where the pairs $\{P_-,u_-\}$ and $\{P_+,u_+\}$ are two distinct algebraic solutions of \eqref{ABIdeal} with \textit{the same} $A$ and $B$. Assuming that $A>0$ (and so $u_x>0$), one must take $\{P_-,u_-\}$ to be the solution with lower pressure and higher speed, while $\{P_+,u_+\}$ must the one with higher pressure and lower speed\footnote{This is required for consistency with the second law of thermodynamics, which demands $\partial_x (su^x)\geq 0$, and thus $s_+ u_+>s_- u_-$ \cite{NovikovThorne1973}.}, namely
\begin{equation}\label{P+P-U+U-}
\begin{cases}
P_-=\frac{A}{3} \left(R-\sqrt{4 R^2{-}3}\right)\, , \\
u_-=\frac{1}{2} \left(2 R^2-R\sqrt{4 R^2{-}3}-1\right)^{-1/2} \, ,\\
\end{cases}\qquad
\begin{cases}
P_+=\frac{A}{3} \left(R+\sqrt{4 R^2-3}\right)\, , \\
u_+=\frac{1}{2}\left(2 R^2+ R \sqrt{4 R^2-3}-1\right)^{-1/2}\, , \\
\end{cases}
\end{equation}
where $R=B/A\in [\sqrt{3}/2,1]$. Finally, one needs to adjust the values of $A$ and $R$ so that $P_-$ and $P_+$ match the required boundary values, i.e. $P_-=P(T_-)$ and $P_+=P(T_+)$.

% \begin{figure}[b!]
%     \centering
% \includegraphics[width=0.45\linewidth]{ShockPP.pdf}
% \includegraphics[width=0.45\linewidth]{Shockuu.pdf}
% \caption{Pressure (left panel) and velocity (right panel) on the two sides of a shock in an ideal ultrarelativistic fluid at zero chemical potential with $A>0$. For each value of $R$, the corresponding point of the blue curve describes the left side of the shock, and the corresponding point of the red curve describes the right side of the shock. See Eq. \eqref{P+P-U+U-} for the analytical expressions.}
%     \label{fig:shockIdeal}
% \end{figure}

\subsubsection{With viscosity}

\begin{figure}[h!]
    \centering
\includegraphics[width=0.45\linewidth]{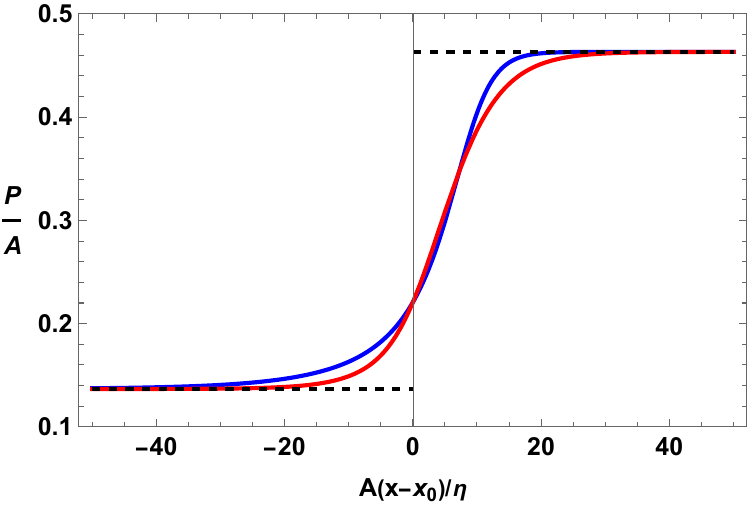}
\includegraphics[width=0.45\linewidth]{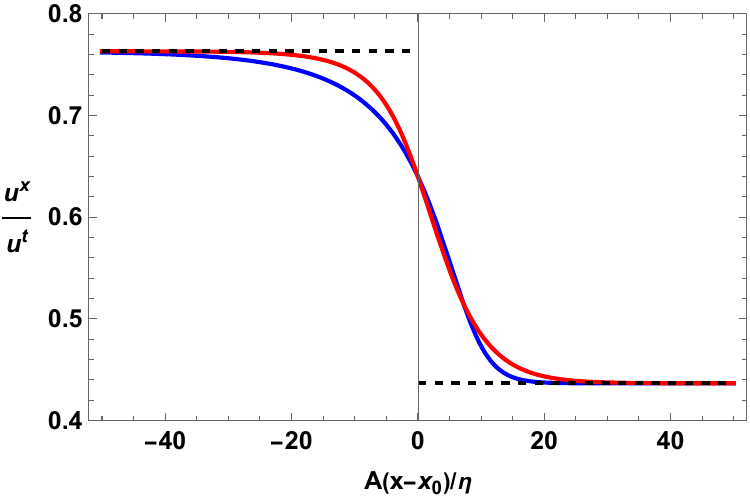}
\caption{Pressure (left panel) and velocity (right panel) across a shock in an ultrarelativistic fluid at zero chemical potential with constant viscosity, and with relaxation time $\tau_\pi =0$ (blue line) and $5\eta/A$ (red line). We have chosen $R=0.9$, and we have set, as initial data, $u(x_0)=(u_++u_-)/2$. The dashed line corresponds to the ideal fluid solution \eqref{discontinuoussolution}.}
\label{fig:SmoothShock}
\end{figure}

In the presence of shear stresses, the conditions \eqref{ABIdeal} generalize to the viscous Rankine-Hugoniot form \cite{OlsonShocks1990pnz,CalzettaShocks2021jpj}:
\begin{equation}\label{ABEckart}
\begin{split}
& 4P \gamma u_x+\pi^{xt}=A \, , \\
&4P u_x^2+P+\pi^{xx}=B \, , \\
\end{split}
\end{equation}
where we are working in the Landau frame.
Using the algebraic constraint $\pi^{x\nu}u_\nu =-\pi^{xt}\gamma+\pi^{xx}u_x=0$, we can rearrange the above system as follows:
\begin{equation}\label{PApixxA}
\begin{split}
\dfrac{P}{A}={}& \dfrac{\gamma{-}Ru_x}{3u_x} \, , \\
\dfrac{\pi^{xx}}{A} ={}& -\dfrac{4}{3} \gamma^2 \left( \dfrac{1{+}4u^2_x}{4\gamma u_x} -R\right)\, .\\
\end{split}
\end{equation}
These equations must be complemented by a relation connecting $\pi^{xx}$ to the gradients. In the framework of Israel-Stewart theory, this relation reads
\begin{equation}\label{longtiudinalm}
\tau_\pi u_x \left[ \partial_x \pi^{xx}-2 \dfrac{u_x}{\gamma^2} \pi^{xx} \partial_x u_x \right]+\pi^{xx}=-\dfrac{4}{3} \eta \gamma^2 \partial_x u_x \, .
\end{equation}
Plugging the second line of \eqref{PApixxA} into \eqref{longtiudinalm}, we obtain a first-order differential equation for $u_x(x)$. Its solutions are a smoothed-out version of the discontinuous shock profile \eqref{P+P-U+U-} \cite{deOliveira:2026abz}, as can be seen in the example provided in Fig. \ref{fig:SmoothShock}.

To have an intuitive understanding of why this is the case, consider the simplified scenario with $\tau_\pi=0$, which corresponds to Landau's first-order theory of relativistic viscosity \cite[\S 136]{landau6}. In this limit, Eq. \eqref{longtiudinalm} reduces to
\begin{equation}
\partial_x u_x =\dfrac{A}{\eta} \left( \dfrac{1{+}4u^2_x}{4\gamma u_x} -R\right) \, .
\end{equation}
Since the right-hand side is proportional to the polynomial $(u_x {-} u_+)(u_x {-} u_-)$, it follows that $u_x(x) = u_+$ and $u_x(x) = u_-$ are exact constant solutions. Consequently, if the initial condition satisfies $u_x(x_0) \in (u_+, u_-)$, the corresponding trajectory $u(x)$ cannot leave this interval, as doing so would require crossing $u_+$ or $u_-$, in contradiction with the standard uniqueness theorems for first-order differential equations. Within this interval, however, one has $\partial_x u_x \sim (u_x {-} u_+)(u_x {-} u_-) < 0$, implying that $u(x)$ decreases monotonically from $u_- $ to $u_+$, approaching each limit asymptotically. With a similar line of reasoning, one finds that, if 
\( u_x(x_0) \notin (u_+, u_-) \), 
the solution inevitably diverges in one direction. The divergence occurs over a lengthscale 
\( \eta/P \), which is typically microscopic \cite{Denicol2012Boltzmann,DenicolANewWay2010xn,Denicol14Momenta2014vaa},
indicating that such a configuration cannot be regarded as physically meaningful.

\subsubsection{But does the fluid really form a shock?}

We have found that, at zero chemical potential, the only time-independent configuration capable of connecting two regions with unequal temperatures, $T_-$ and $T_+$, is a shock profile. We should then ask whether such a solution is indeed the natural mechanism by which the fluid mediates heat transfer between two plates having those temperatures. The answer is almost certainly negative. To begin with, the energy flux appears to propagate from the colder plate toward the hotter one (since $T^{xt}=A>0$ and $P_+>P_-$; see Fig.~\ref{fig:SmoothShock}), in violation of basic thermodynamic reasoning. In addition, the fluid velocity always satisfies $u_- \ge 1/\sqrt{2} \approx 0.71$, implying that the fluid is ``ejected'' by the hotter plate at relativistic speeds, even in the limit where the temperature difference between the plates tends to zero.

If the shock solution is not viable, and if we insist on keeping the stationarity assumption, then the only possible flow profile is one in which the fluid maintains uniform properties throughout the region between the plates. Consequently, the hypothesis that the temperature of the fluid matches that of the plates at the boundaries must be abandoned. This is not entirely unexpected, since the flow velocity already does not coincide with that of the solid surface it is in contact with (given that the fluid penetrates the plates), so there is no reason to expect its temperature to coincide either.

%This naturally leads us to the question of what occurs immediately beneath the plates’ surface. One anticipates the existence of a non-equilibrium transition layer (analogous to the photosphere of a star \cite{Gray2005}) in which the fluid particles experience their first interactions with the material constituents of the plate. Within this layer, the fluid is progressively slowed down, and its temperature continuously relaxes toward that of the plate as it penetrates deeper into the solid. This transition layer can be modeled in detail within relativistic kinetic theory.

\section{Exact evolution in the non-self-interacting limit}\label{freestreaimng}

Here, we discuss the evolution of a boson gas with $\mathcal{C}_{bb}=0$, initialized as in figure \ref{fig:RampandBump} (upper panel). Assuming local thermodynamic equilibrium at the initial time, we have the following initial data:
\begin{equation}
f_0=f_{eq}\left([\varepsilon_0(x)/a]^{1/4}\right) \, \qquad\qquad \text{with} \qquad \varepsilon_0(x)=
\begin{cases}
\varepsilon_\ell, & x\le -L/2\,,\\[6pt]
\dfrac{\varepsilon_\ell{+}\varepsilon_{\mathit r}}{2}+\dfrac{\varepsilon_{\mathit r}{-}\varepsilon_\ell}{L} \, x, & -L/2<x<L/2\,,\\[6pt]
\varepsilon_{\mathit r}, & x\ge L/2\, ,
\end{cases}
\end{equation}
where we recall that $f_{eq}(T)$ is the black-body distribution at temperature $T$.

In the space between the walls ($-L/2<x<L/2$), particles undergo free streaming. Hence, we can write $f(t,x,p^j)=f_0(x-v^x t,p^j)$, where $v^x=p^x/p^0$ is the component of the particle velocity in direction $x$. Note that this formula remains valid also when the location $x-v^x t$ falls inside the walls, because the particle beams emitted by the left wall are always distributed according to $f_{eq}(T_\ell)$ and those emerging from the right wall are always distributed according to $T_{\mathit r}$. Thus, the energy density reads
\begin{equation}
\varepsilon(t,x)=\int \dfrac{g\, d^3p}{(2\pi)^3} \, p^0 f_{eq}\left([\varepsilon_0(x{-}v^x t)/a]^{1/4}\right) \, .
\end{equation}
Expressing the momentum integral in spherical coordinates, we can carry out the integral in the radial momentum analytically, and we are left with
\begin{equation}
\varepsilon(t,x)=\int_{-1}^{+1} \dfrac{dv^x}{2} \, \varepsilon_0(x{-}v^x t) \, .
\end{equation}
This final integral can also be evaluated analytically, taking into account that $\varepsilon_0$ is a piecewise function. The result is
\begin{equation}
\begin{split}
\varepsilon(t,x) & =\dfrac{\varepsilon_\ell}{2t} \left(t-x-\dfrac{L}{2}\right)\Theta\left(t-x-\dfrac{L}{2}\right)+\dfrac{\varepsilon_{\mathit r}}{2t} \left(t+x-\dfrac{L}{2}\right)\Theta\left(t+x-\dfrac{L}{2}\right)\\
&+\dfrac{1}{4t} \left[\varepsilon_\ell+\varepsilon_{\mathit r} +\dfrac{\varepsilon_{\mathit r}-\varepsilon_\ell}{L} \min \left(x+t,\dfrac{L}{2}\right) \right]\min \left(x+t,\dfrac{L}{2}\right)\\
&-\dfrac{1}{4t} \left[\varepsilon_\ell+\varepsilon_{\mathit r} +\dfrac{\varepsilon_{\mathit r}-\varepsilon_\ell}{L} \max \left(x-t,-\dfrac{L}{2}\right) \right]\max \left(x-t,-\dfrac{L}{2}\right) \, ,\\
\end{split}
\end{equation}
which, we recall, is only valid for $-L/2\leq x\leq L/2$. For $x<-L/2$, the particle beams with $v^x>0$ come from deeper inside the wall left, and thus are distributed according to $f_{eq}(T_\ell)$ at all times, while the particle beams with $v^x<0$ are entering the wall from outside, and thus undergo exponential relaxation on a lengthscale $\tau_\ell/v^x$, which further complicates the integral. A similar behavior (but reversed) takes place for $x>L/2$.

\section{Numerical scheme for the relativistic Boltzmann equation}\label{ccccccccc}

We employ a custom relativistic Boltzmann solver for a system of massless bosons with $\nu$ internal degrees of freedom confined to a spatial geometry varying only along the $x$-axis, the kinetic equation reduces to:
\begin{equation}
    p^0 \partial_t f(t, x, \mathbf{p}) + p^x \partial_x f(t, x, \mathbf{p}) = \frac{p \cdot u(t,x)}{\tau_R(t,x)} \left[ f(t, x, \mathbf{p}) - f_{\rm eq}(t, x, \mathbf{p}) \right],
\end{equation}
where $u^\mu = (u^0, u^{x}, 0, 0)$ is the local fluid four-velocity matching the Landau matching condition, and $\tau_R$ is the local relaxation time related to the shear viscosity-to-entropy density ratio by $\tau_R = 5\eta/sT$.

\subsubsection{Momentum Space Reduction and  Discretization}
Because the particles are ultrarelativistic ($E = p = |\mathbf{p}|$) and the geometry possesses azimuthal symmetry around the $x$-axis, the full three-dimensional momentum dependence can be drastically simplified. We integrate out the absolute momentum magnitude analytically by defining an integrated distribution function over the angular variable $\mu = \cos\theta = p^x/p$:
\begin{equation}
    F(t, x, \mu) \equiv \int_{0}^{\infty} p^3 f(t, x, p, \mu) \, dp.
\end{equation}
By performing this integration on the Bose-Einstein equilibrium distribution $f_{\rm eq} = \frac{\nu}{(2\pi)^3} \left[ \exp(p \cdot u / T) - 1 \right]^{-1}$, the analytical local equilibrium state for our integrated variable becomes:
\begin{equation}
    F_{\rm eq}(\mu; T, u^0, u^x) = \frac{\pi \nu T^4}{120 (u^0 - u^x \mu)^4},
\end{equation}
which correctly recovers the standard ultrarelativistic boson energy density $\epsilon = \frac{\pi^2 \nu T^4}{30}$ in the local rest frame. The continuous angular space $\mu \in [-1, 1]$ is then discretized using a Gauss-Legendre quadrature rule of order $N_\mu = 64$. The components of the energy-momentum tensor $T^{\mu\nu}$ are computed at each time step via numerical quadrature
\begin{equation}
\label{eq:Quad}
    T^{00}(t, x) = 2\pi \sum_{j=1}^{N_\mu} w_j F(t, x, \mu_j), \quad 
    T^{0x}(t, x) = 2\pi \sum_{j=1}^{N_\mu} w_j \mu_j F(t, x, \mu_j), \quad 
    T^{xx}(t, x) = 2\pi \sum_{j=1}^{N_\mu} w_j \mu_j^{2} F(t, x, \mu_j),
\end{equation}
where $\mu_j$ and $w_j$ represent the Gauss-Legendre roots and corresponding weights, respectively.

\subsubsection{Space-Time Grid and Boundary Reservoirs}
The spatial domain is discretized on a uniform lattice $x_i = x_{\rm min} + (i - N_b)\Delta x$ for $i \in [0, N_x - 1]$, featuring a total resolution of $N_x = 1024$. We introduce $N_b = 256$ reservoir cells at both boundaries, such that the leftmost cells ($i < N_b$) are coupled to the heat bath at temperature $T_L$, while the rightmost cells ($i \ge N_x - N_b$) are coupled to a heat-bath at temperature $T_R$. Effectively, the $N_x - 2N_b$ cells in the center then represent the physical space $x \in [-L/2, L/2]$.

\subsubsection{Splitting Algorithm and Exact Landau Inversion}
The time evolution proceeds via a multi-stage operator splitting technique, which consists of the following steps
\begin{enumerate}
    \item \textbf{Advection Step:} Spatial transport is solved using a first-order upwind finite-difference scheme, where depending on the direction of the angular velocity $\mu_j$, the intermediate state $F^*_i(\mu_j)$ is advanced using backwards or forwards spatial differences, i.e.
    \begin{equation}
        F^*_i(\mu_j) = \begin{cases} 
        F^n_i(\mu_j) - \frac{\Delta t}{\Delta x} \mu_j \left( F^n_i(\mu_j) - F^n_{i-1}(\mu_j) \right), & \mu_j > 0 \\ 
        F^n_i(\mu_j) - \frac{\Delta t}{\Delta x} \mu_j \left( F^n_{i+1}(\mu_j) - F^n_i(\mu_j) \right), & \mu_j \le 0
        \end{cases}
    \end{equation}
    
    \item \textbf{Landau Matching \& internal relaxation:} Next, to enforce the Landau matching condition $T^{\mu\nu}_* u_\nu = \epsilon u^\mu$ which defines the local fluid four-velocity $u^\mu$ as the timelike eigenvector of the energy-momentum tensor, and the local rest-frame energy density $\epsilon$ as its eigenvalue,  the relevant components of the energy-momentum tensor ($T^{00}_*$, $T^{0x}_*$, and $T^{xx}_*$) of the advected state $F^*$ are calculated using the quadrature rules in Eq.~(\ref{eq:Quad}). Eliminating $\epsilon$ from the system of equations yields a quadratic equation for the fluid velocity $v = u^x/u^0$:
    \begin{equation}
        T^{0x}_* v^2 - \left(T^{00}_* + T^{xx}_*\right)v + T^{0x}_* = 0.
    \end{equation}
    which provides the physical solution
    \begin{equation}
        v = \frac{(T^{00}_* + T^{xx}_*) - \sqrt{(T^{00}_* + T^{xx}_*)^2 - 4(T^{0x}_*)^2}}{2T^{0x}_*}.
    \end{equation}
    from which we obtain the four-velocity components $u^0 = (1-v^2)^{-1/2}$ and $u^x = u^0 v$, as well as the rest-frame energy density $\epsilon = T^{00}_* - v T^{0x}_*$, and the matching temperature $T = (30\epsilon / \pi^2 \nu)^{1/4}$ for the massless boson gas. Subsequently, the distribution is updated via an implicit linear blending using the damping parameter $\alpha = \exp(-\Delta t / \tau_R)$:
    \begin{equation}
        F^{n+1}_i(\mu_j) = F_{\rm eq}(\mu_j; T, u^0, u^x) + \left( F^*_i(\mu_j) - F_{\rm eq}(\mu_j; T, u^0, u^x) \right) \alpha
    \end{equation}
    to describe the internal relaxation dynamics.

    \item \textbf{Reservoir relaxation:} Finally, for coordinates assigned to the reservoir zones, an unconditional relaxation toward the static reservoir equilibrium $F_{\rm eq}(\mu_j; T_{\rm res}, 1, 0)$ is enforced using $\alpha_{\rm res} = \exp(-\Delta t / \tau_{\rm res})$.
\end{enumerate}
If not stated otherwise, in practice, we use a rather conservative time-step $\Delta t = 0.02 \Delta x$.

\bibliography{Biblio}

\label{lastpage}
\end{document}